\documentclass[runningheads]{llncs}
\usepackage{graphicx}
\usepackage{subfig}
\usepackage{xcolor,soul,framed} 
\usepackage{color}
\begin{document}
\title{The Shift to 6G Communications: Vision and Requirements}

\author{Muhammad Waseem Akhtar\inst{1}\and
Syed Ali Hassan\inst{1} \and
Rizwan~Ghaffar\inst{2}\and
Haejoon~Jung\inst{3}\and
Sahil~Garg\inst{4} \and
M.~Shamim~Hossain\inst{5}}

\authorrunning{M.W. Akhtar et al.}

\institute{School of Electrical Engineering and Computer  Science (SEECS), National University of Sciences and Technology (NUST), Islamabad, Pakistan
\email{engr.waseemakhtar@seecs.edu.pk, ali.hassan@seecs.edu.pk}\and
Wi-Fi division of Broadcom, San Jose, CA, USA \\
\email{rizwanghaffar24@gmail.com} \and
Department of Information and Telecommunication Engineering, Incheon National University, Incheon  22012, Korea \\
\email{haejoonjung@inu.ac.kr}\and
Electrical Engineering Department, \'Ecole de technologie sup\'erieure, Montr\'eal, QC H3C 1K3, Canada \\
\email{sahil.garg@ieee.org}\and
Department of Software Engineering, College of Computer and Information Sciences, King Saud University, Riyadh 11543, Saudi Arabia \\
\email{mshossain@ksu.edu.sa}}
\maketitle              % typeset the header of the contribution

\begin{abstract}
The sixth-generation (6G) wireless communication network is expected to integrate the terrestrial, aerial, and maritime communications into a robust network which would be more reliable, fast, and can support a massive number of devices with ultra-low latency requirements. The researchers around the globe are proposing cutting edge technologies such as artificial intelligence (AI)/machine learning (ML), quantum communication/quantum machine learning (QML), blockchain, tera-Hertz and millimeter waves communication,  tactile Internet, non-orthogonal multiple access (NOMA), small cells communication, fog/edge computing, etc., as the key technologies in the \textcolor{black}{realization} of beyond 5G \textcolor{black}{(B5G)} and 6G communications. In this article, we provide a detailed overview of the 6G network dimensions with air interface and associated potential technologies. More specifically, we highlight the use cases and applications of the proposed 6G networks in various dimensions. Furthermore, we also discuss the key performance indicators (KPI) for the B5G/6G network, challenges, and future research opportunities in this domain. 

\keywords{6G \and quantum machine learning\and artificial intelligence\and quantum communication \and block-chain, Beyond 5G .}
\end{abstract}
\section{Introduction}
\label{sec:introduction}
Next-generation communication systems aim to achieve high spectral and energy efficiency, low latency, and massive connectivity because of extensive growth in the number of Internet-of-Things~(IoT) devices. These IoT devices will realize advanced services such as smart traffic, environment monitoring, and control, virtual reality~(VR)/virtual navigation, \textcolor{black}{telemedicine}, digital sensing, high definition~(HD), and full HD video transmission in connected drones and robots. IoT devices are predicted to reach 25 billion by the year 2025~\cite{singh2007study}, and therefore, it is very challenging for the existing multiple access techniques to accommodate such a massive number of devices. Even fifth generation~(5G) communication systems, which are being rolled out in the world at the moment, cannot support such a high number of IoT devices.  Third generation partnership project~(3GPP) is already working on the development of 5G standard and has identified \textcolor{black}{massive machine type communication~(mMTC)}, ultra-reliable and low latency communication~(URLLC), and enhanced mobile broad band~(eMBB) as three main use cases for 5G in its Release 13~(R13)~\cite{holma2016lte}. \par
At the same time, algorithms for the next generation communication systems, which will have the performance higher than that of existing 5G networks, are being developed~\cite{3gppmacro5g}. A typical 5G communication system has the capability to support at most 50,000 IoTs and/or narrowband IoT~(NB-IoT) devices per cell~\cite{holma2016lte}. Specifically, a more robust network must be designed to realize the massive access in beyond 5G~(B5G)/6G communication systems. We now discuss comprehensive literature that has appeared on various dimensions of 6G networks.\par

\subsection{Vision and Literature Survey}
 Currently, there is little information about the standards of 6G. However, it is estimated that the international standardization bodies will sort out the standards for 6G by the year 2030~\cite{6gtime1}. \textcolor{black}{The work at some of the research centers has shown that 6G will be capable of transmitting a signal at a human computational capability by the year 2035~\cite{6gera}.} 
%===========
%Fig 1
%===========
\begin{figure*}[!b]
\centerline{\includegraphics[width=\textwidth]{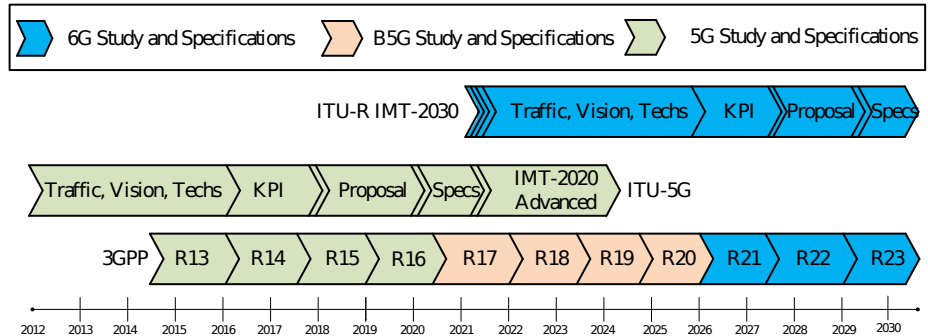}}
\caption{\textcolor{black}{A tentative timeline of standards development for 5G, B5G, and 6G \cite{6gtime1,6gtime2}.}}
\label{6Gstanddards}
\end{figure*}
 While the rollout of 5G is still underway, the researchers across the world have started working to bring a new generation of wireless networks. A tentative timeline for the implementation of 5G, B5G, and 6G standards by international standardisation bodies is shown in Fig.~\ref{6Gstanddards} with respect to the vision of 6G wireless networks. International Telecommunication Union Radiocommunication sector ~(ITU-R) issued the requirements of International Mobile Telecommunications-2020 (IMT-2020 Standard) in 2015 for the 5G network standards. At the same time, 3GPP issued R13 for 5G standards. It is predicted that ITU will complete the standardization of 6G~(ITU-R IMT-2030) by the end of the year 2030, whereas 3GPP will finalize its standardization of 6G in R23~\cite{6gera}. ITU has established a focus workgroup for exploring the system technologies for B5G/6G systems in July 2018~\cite{ITU_2030ref2}. The Academy of Finland has founded, 6Genesis, a flagship program focusing on 6G technologies, in 2018~\cite{finland6Gref3}. Similarly, China, the United States of America, South Korea, Japan, Russia have also started the research for B5G/6G communication technologies~\cite{6g1,6gera,6gR1,6gR0,b5g01}. \par
 The vision of 5G technologies is extended for the 6G networks by speculating the visionary technologies for next-generation wireless systems in \cite{6gera}. Different networking scenarios are presented in~\cite{li2009future,zhang20196g,gawas2015overview,basar2019wireless}. \textcolor{black}{The authors in \cite{li2009future} and \cite{zhang20196g} give a predictive technical framework for industries in future generations of communication systems mainly focusing on the specifications of future generations of communication system.} Cell-less architecture, decentralized networking, and resource allocation, and three-dimensional radio connectivity including the vertical direction are expected in next-generation communication systems. The evolution of wireless systems from 1G to 6G is outlined in~\cite{gawas2015overview}. The authors in~\cite{basar2019wireless} presented the role of intelligent surfaces in the architecture of 6G networks. \par
The potential role of artificial intelligence~(AI) in future communication networks is discussed in \cite{letaief2019roadmap}. The authors in~\cite{saad2019vision,yajun20196g}, have focused on the vision for the next generation of wireless communication systems. Blockchain and AI are the potential technologies for the next generation communication systems. Blockchain can be used for efficient resource sharing and AI can be implemented for the robust, self-organizing, self-healing, and self-optimizing wireless network~\cite{8726067}.  \par
 
 By using millimeter-wave (mmWave) and terahertz~(THz) frequency bands, massive bandwidth, and highly directive antennas will be available to the 6G mobile devices to enable new applications and seamless coverage \cite{9163026}. Federal Communications Commission~(FCC) has commercialized these frequency bands in 2019~\cite{3gppNR}. Ultra-high-precise positioning will become available with 6G due to high-end imaging and direction-finding sensors, just like human eyes and ears. 6G mobile phones could be equipped with capable robots and intelligent algorithms~\cite{6gera}.\par 
 %===========
%Fig 2
%===========
\begin{figure*}[t]
\centering
\centerline{\includegraphics[width=\textwidth]{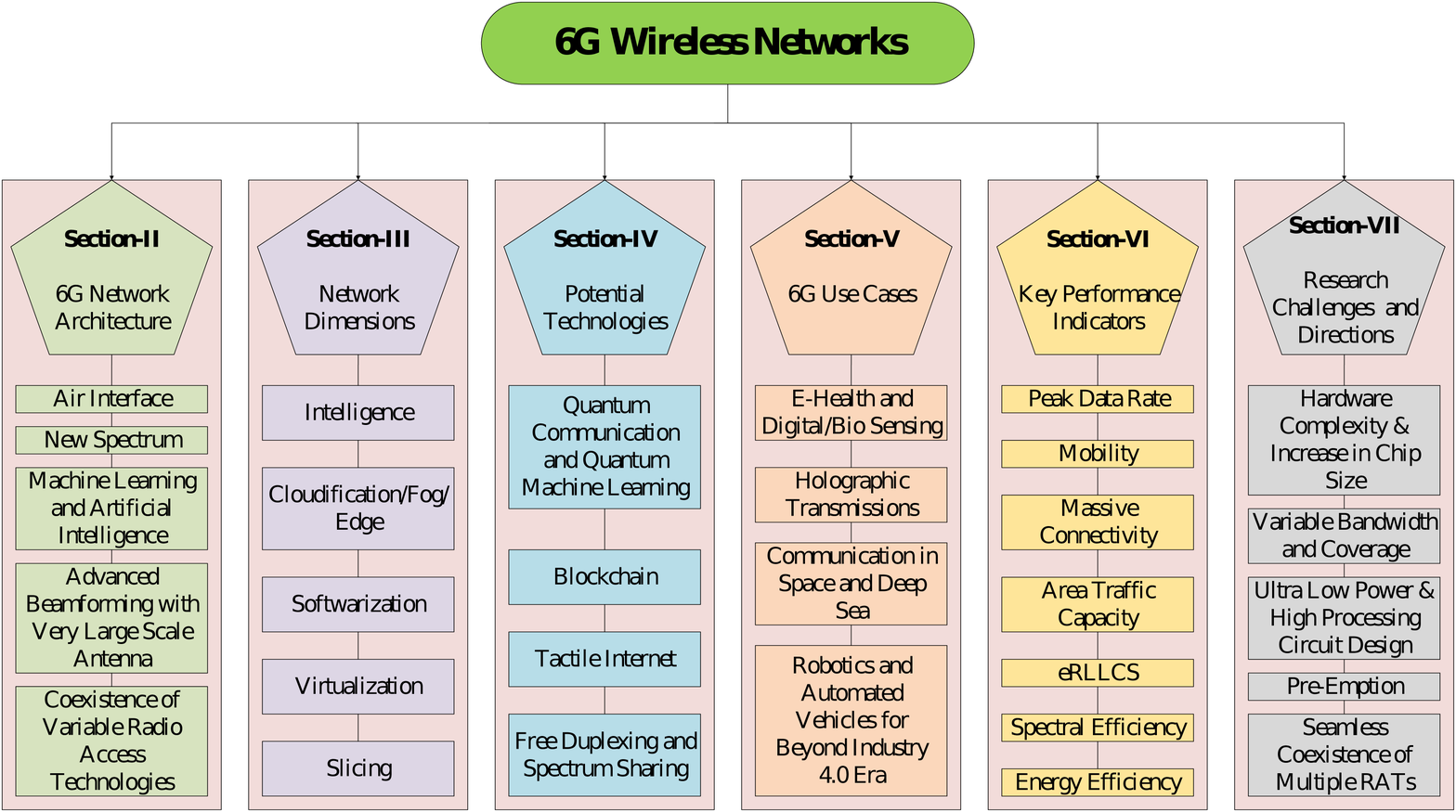}}
\caption{Taxonomy of the paper.}
\label{taxanomy}
\end{figure*}
The latency of the network in 6G will be minimized by using super-fast and high computational power processors both at the network and end devices. The mobile phone of the future network will be intelligent enough to sense the environment and give the precautionary and preventive measures. For example, these mobile phones will be capable to detect the air pollution level, toxic food materials, and explosive materials around us. \textcolor{black}{These phones will replace the wallet, hard cash, and wristwatches by providing digital currencies, and smartwatches, respectively. Similarly, smart goggles will replace glasses and smartphones.} It is anticipated that 6G cellphones, coupled with the incredibly high directive and beam-steering antennas, would be capable enough to see through the walls by reconstructing the images by receiving the signals from multiple levels of density of the environment in the vicinity~\cite{6g1}. This feature would be useful for extracting minerals and elements from rocks, exploring underground natural reserves, and detecting arms. Apart from this, 6G mobile phones will have tremendous features of providing position, location, and range with very high accuracy. This will be helpful for maritime and underwater communication and positioning. \par

Self-driving cars, which are already being developed in the initial phase these days, would make human life safer and more comfortable~\cite{6gera}. Holographic technologies and VR~/~augmented reality~(AR) will break the barrier of distances. The \textcolor{black}{digital} revolution has transformed the way we play, talk, or work. In the recent era of the digital \textcolor{black}{revolution}, 5G has become the center of attention for everyone. Soon the mobile devices in our pockets will get the wireless speed approaching the fiber optic transmission speed, bringing 3D imagery, television, online games, and many other applications that we never imagined into our tablets or mobiles.\par

Special attention is paid to the improvement of the traffic prediction in \cite{liu2018mobile}. Following the 6G vision and service requirements, some use case scenarios for the 6G, such as autonomous vehicles, smart cities, flying networks, holographic, \textcolor{black}{telemedicine}, and Tactile  Internet, are discussed in \cite{3}. Moreover, the reliability of the future wireless network is forecasted to be at the same or higher level as that of today's wired communication networks. \par
Some potential key enabling technologies encompassing blockchain-aided decentralization of the network and machine learning~(ML)-based intelligent communication system for the 6G are discussed in \cite{6}. A comparative analysis between the key performance indicators~(KPIs) for 5G and 6G is carried out in \cite{7}. Practical applications including holography, ML, VR, Internet-of-Things, visible light communication~(VLC), automated driving is discussed in\cite{8}. \par
\subsection{Contribution}
The objective of this article is to draw a complete picture of “how the 6G will look like?”.  We cover different dimensions and aspects of 6G focusing on the projected 6G system architecture, potential technologies, network dimensions, KPIs, applications, and use cases. \textcolor{black}{The taxonomy of the paper is shown in Fig.~\ref{taxanomy}, which gives a pictorial view of all sections and subsection presented in this paper.} The contributions of this article are summarized as follows.\par 
\begin{itemize}
\item {} We discuss in detail about the projected 6G system architecture. We highlight all the essential network elements of 6G system architecture and discuss the new air interface development, application of AI and ML, utilization of new spectrum, the coexistence of variable radio access technologies~(RATs), and intelligent and smart beamforming, etc. 
\item {} 6G network dimensions that include, cloudification/ fog/ edge computing, intelligence, softwarization, and slicing are discussed in detail. 
\item {} Tactile Internet is one of the main use cases of 6G networks and holographic verticals are the potential application. Space tourism and communication in space, handled by 6G networks, is discussed. 
\item {} Towards the end, we discuss multiple challenges and research directions that include the increase in chip size, beamforming for mobile users, pre-emptive scheduling, low latency, and high reliability, and variable bandwidth and coverage issues that will be faced due to the coexistence of all networks.
\end{itemize}
The remaining paper is outlined as follows. The next-generation wireless networks system architecture is described in Section II. Next-generation network dimensions are given in Section III. Some prominent use cases and applications of 6G communication systems are presented in Section IV, followed by potential key enabling technologies for 6G are given in Section V. Section VI describes KPIs for 6G. Research challenges and future directions are given in Section VII and finally, the paper is concluded in Section VIII.

\section{6G System Architecture}
Next-generation wireless networks will consist of massive connected devices and with the base stations~(BSs)/ access points~(APs) leading to mMTC. 
Multiple BSs/APs may serve one or more devices simultaneously to form a coordinated multi-point~(CoMP) transmission~\cite{comp1}. The huge amount of data produced by massive devices will require very high-performance processing units and robust backhauling links. The central processing units may utilize ML and AI algorithms and the backhauling links may utilize optical fiber and or photonic communications. \textcolor{black}{Remote user, in 6G communication systems, can use several relays or transmitters for a remote user to transmit, and the user's SINR may be improved by using the technique of diversity as in virtual MIMO systems.} 

\begin{figure}[t]
\centerline{\includegraphics[width=0.45\textwidth]{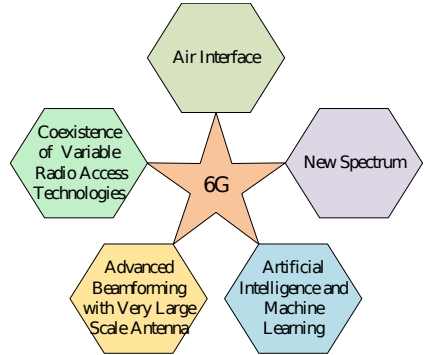}}
\caption{6G wireless network elements.}
\label{6G}
\end{figure}

\textcolor{black}{By intelligent networking, all the end devices would be aware of the location and features of BSs/APs in their vicinity, and all of the BSs/APs would be aware of the locations, features, and QoS requirements of devices in their vicinity.} \textcolor{black}{Robust interference management / optimization techniques can be applied to maximize the efficiency of the wireless network.} Central processing units will be fast enough to manage and switch the resources~(bandwidth, time, power) among multiple end-users, and data processing will be conducted at the base-band processing units~(BPUs). %===========
Fig.~\ref{6G} depicts some of the major components in the 6G system architecture, that will cause a major paradigm shift towards the realization of 6G standards. The air interface is the main component that causes a major improvement in the wireless generations. Orthogonal frequency division multiplexing~(OFDM) played a major role in the development of 4G, as code division multiple access~(CDMA) was the key player in 3G. Similarly, the development of the new air interface will be an essential component of 6G system architecture.\par

AI and ML is another crucial component of the 6G system architecture. AI and ML will play an important role in the self-organization, self-healing, self-configuration of 6G wireless systems. Spectrum congestion has also pushed the 6G to adopt a new spectrum for communication. Therefore, this new spectrum will also be an active component in the 6G system architecture. Since 6G will accommodate a wide range of communication devices ranging from IoTs to live HD video transmission, 6G will need to be in line with all previous technologies. Therefore, a flexible and multi-radio access technologies~(RAT) system architecture will be an essential component in the 6G network. 
\subsection{Air Interface}
Since 6G will concentrate on the current terahertz frequency range with extremely wide bandwidths available, it will bring up new obstacles to interact efficiently at these frequencies. Getting a secure transmission infrastructure that has an adequate range and isn't power-hungry will be the answer here. The availability of incredibly wide bandwidths would change the emphasis from spectrally optimized solutions to improved coverage solutions. In these new frequency spectrums, the tradeoff between spectrum performance, power efficiency, and coverage will play a key role in developing devices. This will lead to the design of a modern air interface where more consideration can be paid to single-carrier systems. The OFDM scheme would be revisited for lower frequency ranges where spectral efficiency will be important as it does not use the energy effectively because of the cyclic prefix, which is just the duplication of information and does not hold any additional information. Furthermore, a high peak-to-average power ratio~(PAPR) makes the power amplifiers complex and expensive. \par
\iffalse
The receivers, in 6G, now can do some complex algorithms to get rid of inter-symbol-interference~(ISI)~\cite{airint2,airint3}. Therefore, 6G should adopt a better air interface to meet the requirement of 2030's networks. 
\fi

Many researchers have proposed the non-orthogonal multiple access~(NOMA) as a promising new scheme for the B5G/6G mobile networks~\cite{ZDCNOMA,noma1,nomamec1}. In NOMA, all of the users are allowed to access the complete resource~(frequency band) simultaneously. Some researchers have suggested the rate-splitting multiple access~(RSMA) as a new access technology for 6G communication systems~\cite{rsma1,rsma2,rsma3}. Both NOMA and RSMA rely on the successive interference cancellations~(SIC) to decode the information for the user. RSMA uses the SICs to decode the common message firstly and then decode the private message. Both schemes need to be matured enough before practical deployment. \textcolor{black}{A new AI-based software-defined air interface is presented in ~\cite{airint1}, where the authors proposed an intelligent air interface switching system for user QoS enhancement.}  \par 
3GPP release 15, reveals the specifications for the 5G-New Radios~(NR), in which multiple waveform configurations and two sets of frequencies are defined. By adopting the variable numerologies~(symbol duration, sub-carrier spacing, and pilot spacing), we can give the transmitter leverage to self-organize and self-configure according to the channel conditions and service required. \textcolor{black}{This is often useful on different measurements. For instance, by reducing the symbol length, low latency can be achieved, and increasing the spacing of the sub-carrier can be helpful in reducing the phase noise in mmWave and sub-mmWave. In high mobility situations, optimizing the sub-carrier width can also be helpful for Doppler shift compensation.} \par

\subsection{New Spectrum} 
mmWave is already a candidate for 5G, but it is not exploited to its full potential as the beamforming algorithms are not mature enough. \textcolor{black}{It requires improvements in the networks when personal BSs and satellite connectivity can get merged into cellular communication.} In the previous generations, the spectrum is divided for multiple services, for instance, television~(TV) services, military communications, and cellular communication~\cite{mmwave1}.\par
Therefore, the idea of using an unlicensed spectrum is proposed, i.e., to use the mmWave, THz band, and visible light spectrum, simultaneously~\cite{mmwave2,mmwave3,mmwave4,vlc1,vlc2}. These bands are never used for any communication. \textcolor{black}{The problem with the higher frequency band, though, is that the signal is attenuated very rapidly with regard to the distance travelled.} For example, a 3G or 4G BS can have a coverage of about several miles whereas a 5G or 6G BS coverage may limit to only a few hundreds of meters. To resolve this issue in mmWave and THz communications, the idea of using massive multiple inputs and multiple outputs~(MIMO) and beamforming emerged, which is described in the next subsections.
%%%%%%%%%%%%
%Fig 3
%===========
%===========
%Fig 4
%===========
 \begin{figure*}[t]
\centerline{\includegraphics[width=\textwidth]{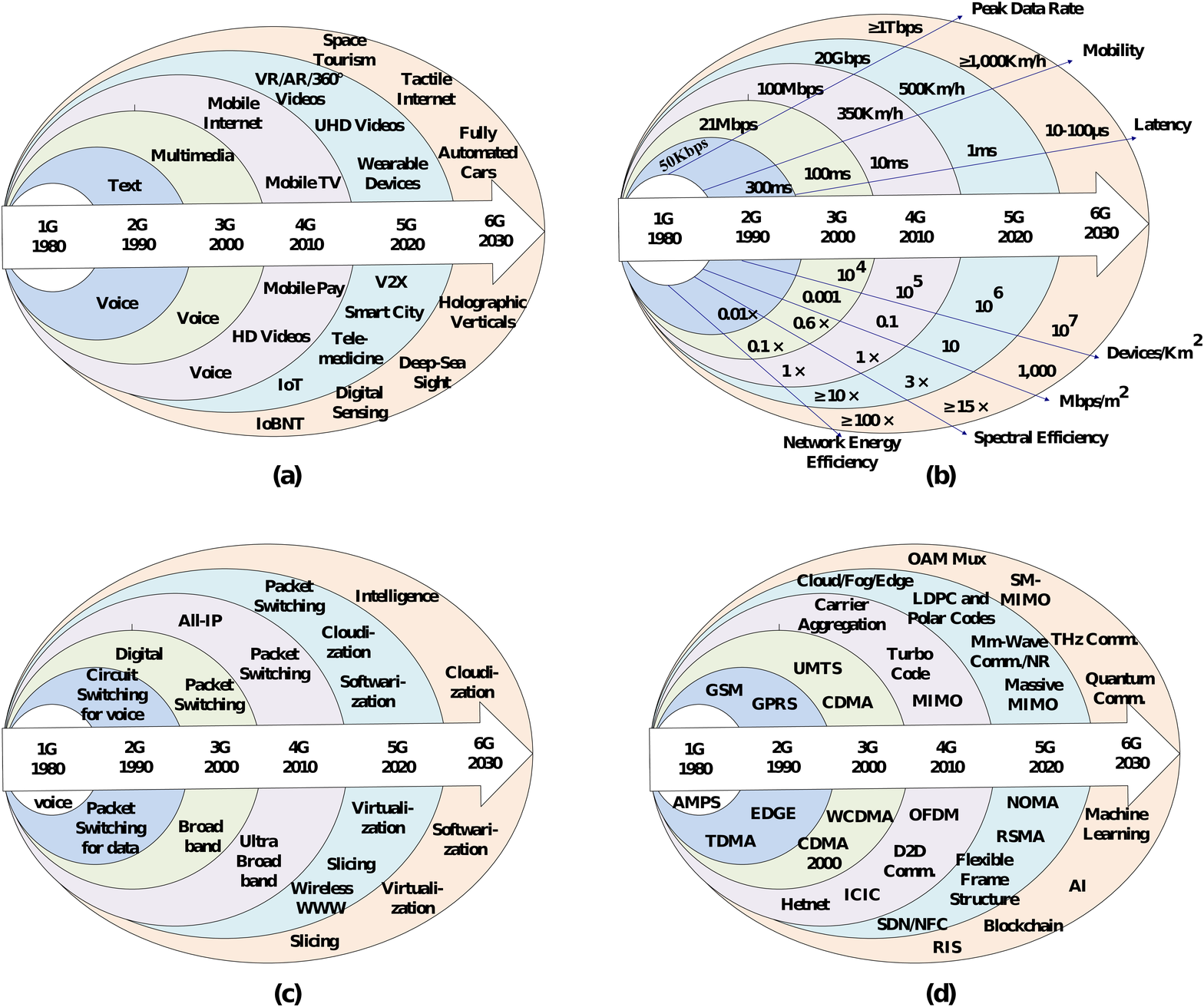}}
\caption{Evolution of wireless communication, with timeline, from 1G to 6G based on (a) Applications (b) KPIs (c) Network characteristics and (d) Technological development}
\label{ techn_comp }
\end{figure*}
\subsection{Artificial Intelligence/Machine Learning} 

\textcolor{black}{By offering pervasive, secure, and close proximity-instant wireless networking for humans and computers, 6G wireless communication networks would be the core of society's digital transition. 
A broad variety of emerging developments, such as self-driving cars and voice assistants, have been made possible by recent advancements in ML research. B5G/6G wireless networks have increased complexity, requiring smarter methods for handling any losses and handling network features, detecting anomalies, and understanding KPI trends. This can be done by introducing solutions for ML and SDN. In order to preserve a certain level of KPI, ML/AI will boost the decision-making process. The operation and implementation of RAN for 6G needs a new strategy. Incorporating AI in wireless algorithms (e.g., for channel estimation, for channel state information~(CSI) feedback, and decoding, etc.) may bring a change in the direction of these algorithms~\cite{aiml1}.}
Application of ML, DL \cite{7494596}, and AI algorithms to the communication network, we can instantly manage the resources as per the user requirements. \textcolor{black}{The probability of choosing the best solution is improved in this way and the network can maintain its optimum state.}

\iffalse
It helped people to eliminate the limitations of physical capabilities as we thought of the industrial revolution, and as we thought of the digital revolution, it helped corporations to eradicate the limitations of human capabilities. Network operations generally work on the principle of simplicity and consistency. \textcolor{black}{Because of huge numbers of connected devices, the network receives more and more data, and the networks are being overloaded.}\textcolor {black}{Wireless networking algorithms have now achieved a stage of complexity where it is challenging to accomplish more advancement.} 
\textcolor{black}{A telecommunication network can easily be compared to the vehicular traffic network in big cities both in terms of complexity and traffic volume~\cite{aiml2}. They both can also be managed similarly. The flow of data packets in the telecommunication network is similar to cars entering the cities both on streets and highways. However, the movement is not smooth. There is a situation when traffic is concentrated in sensitive places. The usual operation for maintaining the network cannot accommodate such a large amount of incoming data. \textcolor{black}{Typically, the appropriate emergency protocol is followed anytime such an event arises.} In a short time, even a highly skilled operator cannot limit the data pressure, which may even make the problem worse.}\par 
\fi

\subsection{Advanced Beamforming with Very Large Scale Antenna~(VLSA)}
The idea of beamforming is to steer the beam to only the desired direction or user. \textcolor{black}{Since energy is not spread in all directions, the transmission range is thus improved by concentrating the beam in one direction.} 
\subsubsection{Intelligent Reflecting Surfaces~(IRSs)}
Intelligent Reflecting Surfaces~(IRSs) can be the potential area for beamforming in 6G~\cite{ris1}. IRSs are composed of thin electromagnetic materials, which can reflect/configure the incoming electromagnetic rays in an intelligent way by configuring the phase of reflected rays by a software~\cite{ris1}. Indeed, IRSs use at a large number of low-power and low-cost passive elements to reflect the incident signals with configurable phase shifts without the requirements of additional power, encoding, decoding, modulation, demodulation requirements. IRSs are installed on the important points and locations such as high-rise buildings, advertising panels, vehicles~ (cars, airplanes, unmanned aerial vehicles (UAVs)), and even the clothes of the pedestrians. The main advantage of the IRS is that it can enhance the signal-to-interference-plus-noise-ratio~(SINR) with no change in the infrastructure or the hardware of the communication network. Also, there is no need for extra power required for the installation.\par 
IRS can reduce the hardware complexity at the receiver and the transmitter by reducing the number of antennae installed at them, thereby, reducing the radio frequency~(RF) chains at the transmitter and the receiver. IRS can replace the conventional relays system due to its advantages in terms of power, spectral efficiency, and reduced hardware complexity~\cite{ris5}. IRS can be used in the deep-fade and non-line-of-sight~(NLOS) communication environment. The principle by which SINR is enhanced at the receiver is optimally controlling the phases of the incident ray at multiple elements of the IRS, to produce useful beamforming at the receiver~\cite{ris5}. Degradation factors such as noise and interference have no impact on the IRS. All these features of the IRS make it a promising technique for the B5G/6G communication systems. \par
%=========iffalse=====================
\iffalse
There are hundreds of antennas at the BS and twenties of the antenna on the mobile. At the higher frequencies, the size of the antenna is decreased. Hence, there is enough room in the mobile station and the BS to place multiple antennas. This is a new communication paradigm. However, the challenge with the beamforming is that with the mobility of the mobile user, or slightly moving the direction of the mobile, there will be a huge variation in the beamforming.
\fi
%==========fi======================
\subsubsection{Orbital Angular Momentum~(OAM)-aided MIMO}
A new dimensional property of the electromagnetic waves~(EW) was discovered in the 1990s termed as the orbital angular momentum~(OAM). This discovery promised the transmission of multiple data streams over the same spatial channel. An EM wave carrying the OAM has the phase rotation factor of $\exp(-jl\phi)$, where $l$ is OAM state number represented in integer and $\phi$ is transverse azimuth angle~\cite{oam1,oam2,oam3}. The main advantage of OAM over other beamforming techniques is that OAM can have an unlimited number of orthogonal modes, which allows the EW to multiplex multiple data streams over the same spatial channel, thereby, enhancing the spectral efficiency and transmission capacity. OAM support a high number of user in mode division multiple access~(MDMA) scheme without utilizing extra resources~(i.e., frequency, time, and power). The flexibility of OAM to be used in narrowband and wideband frequency hopping scheme makes it an attractive scheme for low probability of interception~(LPI) applications. OAM-based MIMO systems have advantages over the conventional MIMO systems in terms of capacity and long-distance line-of-sight~(LoS) coverage~\cite{oam4}. Therefore, OAM has great potential for applications in 6G wireless networks.

\subsection{Coexistence of Variable Radio Access Technologies} 
\iffalse
6G should lead to a ubiquitous communication system where the choice of best communication network will not be left to the users.
\fi
\textcolor{black}{6G can lead to a ubiquitous networking infrastructure where users would not only be left with the option of selecting the best communication network. Each node in this network would, however, be intelligent enough to sense the conditions of the channel and the specifications of QoS at any other node.} For example, the use case and the availability of network will decide the network as cellular, wireless LAN, Bluetooth, and ultra-wideband (UWB), etc. 6G communication standard should, therefore, be designed in such a way that it will converge all of the wireless technologies. \textcolor{black}{Communication with Wi-Fi, Bluetooth, UWB, VLC, UAVs, biosensors, and satellite communications can all integrate into 6G and should fall under one standard such that all of them can connect with each other.} The Wi-Fi operating at 2.4 GHz has already entered deeply into IoTs as most of the appliances are now connected through this network~\cite{coex1,coex2,coex3}.\par
By merging all these technologies, 6G would be able to utilize the massive infrastructure deployed by previous technologies, which otherwise can cost 6G a huge revenue. The features in the previous technologies, such as network densification, high spectral efficiency, high throughput, low latency, high reliability, and massive connectivity should be converged in 6G. 6G technology should also keep the trend of offering new services by applying the new technologies, such as AI/ML, VLC, quantum communications~(QC), and quantum machine learning~(QML). These services may include but are not limited to smart cars, smart homes, smart wearable, and 3D mapping~\cite{coex4}. \par
Fig.~\ref{ techn_comp } gives an overview of the evolution of the wireless generation, with timelines, from 1G to 6G with respect to applications, KPIs, network characteristics, and technology. Fig.~\ref{ techn_comp }(a) shows that a major leap in the application is observed with 4G. 4G introduced mobile Internet, mobile TV, and HD videos. AR/ VR, ultra-HD~(UHD) videos, wearable devices, vehicle-to-infrastructure~(V2X), smart city, telemedicine, and IoTs concepts are introduced in 5G. 6G is projected to have applications such as space tourism, Tactile Internet, fully automated cars, holographic verticals, deep-sea sight, digital sensing, and Internet-of-bio-Nano-things~(IoBNT). Fig.~\ref{ techn_comp }(b) shows that how KPIs are changing with the evolution of wireless generations from 1G to 6G.  Fig.~\ref{ techn_comp }(c) shows the evolution of the network characteristics with wireless generations. All Internet protocol~(IP) and the ultra-broadband concept is introduced in 4G. The concepts of cloudification, softwarization, slicing, virtualization, and wireless worldwide web~(WWW) are introduced in 5G. Integration of intelligence with cloudification, softwarization, slicing, and virtualization will be introduced in 6G communication systems. Fig.~\ref{ techn_comp }(d) depicts the evolution of technologies with the development of wireless communication generations. The initial stage of the wireless communication system is the development of the advanced mobile phone system~(AMPS). Global systems for mobile~(GSM) and general packet radio systems~(GPRS) family is developed in 2G wireless systems. Code-division multiple access~(CDMA) family shifted the wireless systems from 2G to the 3G. OFDM with the integration of turbo codes and MIMO systems is the key technology for 4G communication systems. 5G communication systems brought some new technologies such as cloud/fog/edge computing, massive MIMO, SDN, mmWave and sub mmWave~(NR) along with low-density parity-check~(LDPC) and polar codes. ML, AI, blockchain, THz communication, orbital angular momentum multiplexing~(OAM Mux), spatial Modulation~(SM)-MIMO and intelligent re-configurable reflecting surfaces are the new technological domains in 6G. \par
%===========
%Fig 5
%===========
\begin{figure*}[t]
\centerline{\includegraphics[width=\textwidth]{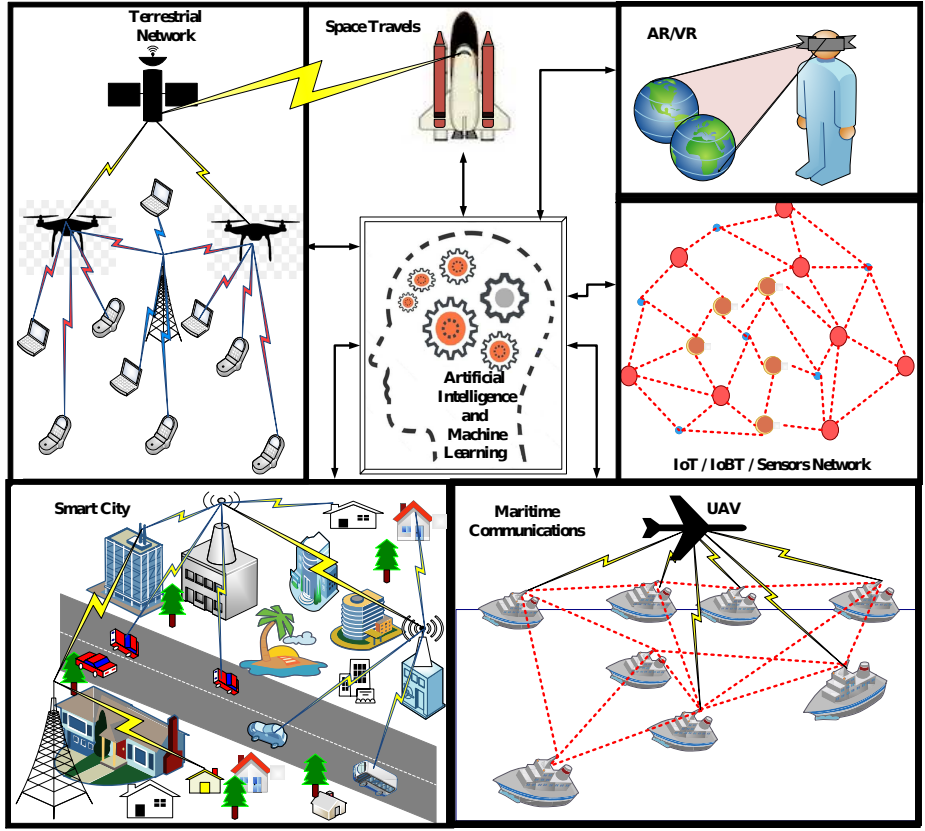}}
\caption{\textcolor{black}{A depiction of space-air-ground-sea based integrated next-generation communication system with a wide range of applications}.}
\label{6GNW2}
\end{figure*}
\section{Network Dimensions}
In this section, we give an overview of the network dimensions of 6G networks. Network intelligence will be an essential component of 6G networks and the network will take actions dynamically according to the environmental conditions. \textcolor{black}{The idea of clouds, fog, and edge computing is applied for fast access to services.} The features of self-optimization, self-organization, self-reconfiguration will be achieved through softwarization, virtualization, and slicing. The detailed discussion on each network dimension is given as follows.
\subsection{Intelligence}
Researchers believe that AI will play a defining role in the IoTs and IoBNTs driven world \cite{8884231}. The potential shift from 5G to the 6G will be to determine an efficient way to transmit data.  The ideal system will be the one that is free from human intervention at all~\cite{aiml4}.
\subsection{Cloudification/Fog/Edge}
Thousands of sensors are installed in the industries and hundreds of the sensors are installed in homes. It is very difficult to connect all these sensors with wires~\cite{cloud1}, and, all these devices can produce a large amount of data. Also, these devices are smart and intelligent, capable of making smart decisions and less processing power. Therefore, we need to offload the data from cloud to edge and device end. To reduce the processing delay, we need to shift the process near to end devices in terms of cloud/fog. We need to place the workload closer to the edge for a better quality of service.

\subsection{Softwarization} Main driving force behind the development of B5G and 6G networks is to provide services such as self-organization, configurability, programmability, flexibility, and heterogeneous use-cases. It is difficult to install the hardware equipment which provides all of the mentioned functionalities. By realizing the functionalities by underlying networks, softwarization and virtualization have emerged as the two most demanding paradigms for B5G/6G networks~\cite{cloud5}.\par 
Softwarization is the term used for the set of interfaces and protocols which can allow the network to be configured in software by decoupling the control and user plane. The user plan usually consists of a set of distributed and stateless routing tables at which packet switching is performed at a very high speed. These tables are updated by the centralized control plan which maintains the end-to-end routing information for multiple services. Data and control management operations are exchanged between the service consumer and the SDN provider~\cite{soft1}. SDN provider ultimately forwards the required service to the service consumer. These services are controlled by the service consumer by taking acting on these virtual resources.
\subsection{Virtualization} Network function virtualization enables the software functions to be performed in the virtual machines and allows the access of common shared physical resources such as storage, networking, and computations. Containers are used to instantiate multiple functions within the same virtual machine. Dynamically varying network demand such as offered services and network traffic can be handled by dynamically instantiating the virtual machines. \par
The services, which can be virtualized, include but are not limited to load management, mobility management, baseband processing unit services, evolved packet core~(EPC) functions. Network function virtualization~(NFV), unlike traditional mobile networks, provides the leverage to route the packet of each service between virtual network functions~(VNFs)~\cite{slice1}. Further, the routing services are provided with very few overheads. Also, routing and traffic flow are smoothly maintained without any interruption even if a new VNF is added or removed according to the traffic demands. 

\subsection{Slicing} One of the key network abilities that will allow us to build a flexible network on top of the common physical infrastructure is network slicing. As 5G continues to take shape, network slicing will become the fundamental technology to enable a wide range of use cases. Taking a single piece of network infrastructure and being able to cost-effectively deliver multiple logical networks over the same common physical infrastructure~\cite{slice1}. The slices can be allocated to some specific use-cases such that we can have a slice for IoTs, slices can be allocated to a class of service, slices can be allocated to a class of customers, slices can be allocated to some specific mobile network operators, slices can also be allocated to network types such as wireless vs wired or consumer vs businesses.\par
%%%%%%%%%%%%
%Fig 6
%===========
\begin{figure}[t]
\centerline{\includegraphics[width=\textwidth]{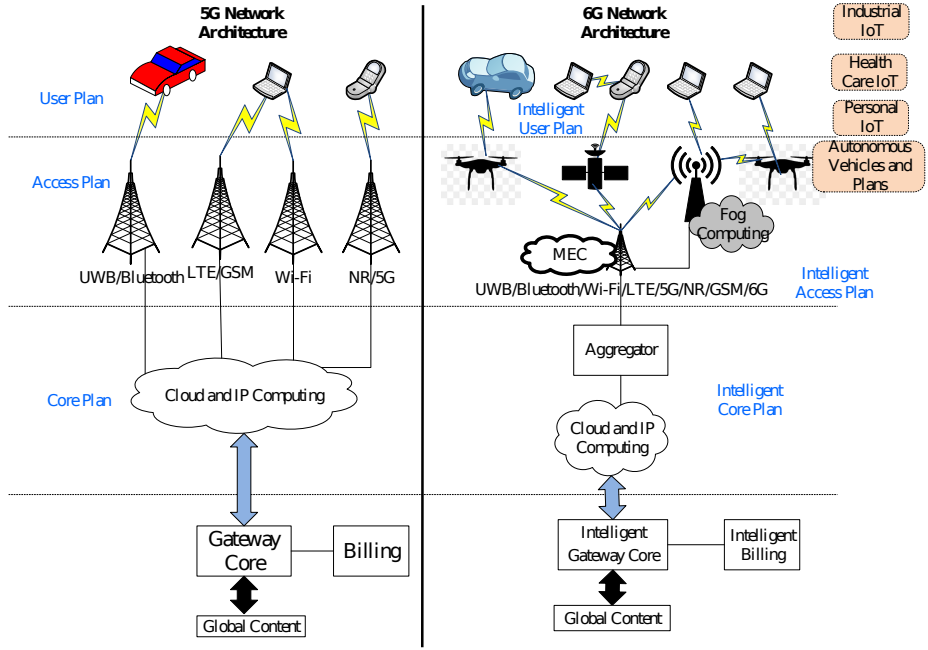}}
\caption{\textcolor{black}{A comparative analysis between 5G and 6G network architecture.}}
\label{6garch}
\end{figure}
\textcolor{black}{In network slicing, the biggest difficulty is configuring new slices, since it affects all the network components. However, with the need to create customized services and deliver a service with a very specific requirement, we can create slices such as slice for automotive, healthcare, utility.}  \par
Fig.~\ref{6GNW2} gives a pictorial overview of the 6G wireless network that covers all aerial-ground-sea communications. As shown in Fig.~\ref{6GNW2}, 6G will make it possible to communicate with the devices with very low data-rates such as biosensor and IoTs, and at the same time, it will enable high data rate communication such as HD video transmission in smart cities. Communication will be possible in a fast-moving bullet train, airplane. It also shows that all of the networks will be merged all together. \textcolor{black}{Further, the buildings and surfaces in smart cities can be equipped with the IRS that could enhance the coverage and quality of service (QoS) of each communicating device. For the maritime communication scenarios, the robust underwater data links will enable the communication between ships, submarines, and sensors at the deep sea level \cite{seadrone,seadrone1}. Besides, innovative technologies such as AR/VR, haptics, and ML will further reduce the effect of physical distances around the globe.} \par
\textcolor{black}{Fig.~\ref{6garch} depicts a comparative analysis of the network architecture of 5G and 6G. The 6G core network is shown to have upgraded to the basic 5G core network on the basis of intelligence, high computational power, and high capacity. By integrating BSs/APs, satellites, and UAVs, the access network is upgraded in a similar way. There is a vertical hand-off in 6G in addition to the horizontal as in that of 5G. In addition, fog computing and MEC are an integral component of the 6G network infrastructure, that reduce latency and bandwidth utilization for regularly needed services by a massive number of devices on the user plan.}

\section {Potential Technologies}
Based on the vision of 6G and its network architecture, we now elaborate on the key enabling technologies for 6G wireless networks in this section. Various state-of-the-art technologies must be utilized together to enable the key features of 6G.
\subsection{Quantum Communication and Quantum ML}
Quantum technology uses the properties of quantum mechanics, such as the interaction of molecules, atoms, and even photons and electrons, to create devices and systems such as ultra-accurate clocks, medical imaging, and quantum computers. However, the full potential remains to be explored. A quantum Internet is a way of connecting quantum computers, simulators, and sensors via quantum networks and distributing information and resources securely worldwide~\cite{qml1}.\par
In October 2018, the European Commission launched the Quantum Flagship, a 1 billion Euro project for over ten years involving 5000 scientists to support quantum research in the EU with the goal of creating a quantum Internet~\cite{qml1}. For the next decade, the EU plans to develop and deploy a secure pan-European QC infrastructure (QCI) to become the backbone of a quantum Internet.  QCI will make quantum cryptography a part of conventional communication networks, protecting the EU sensitive data and digital infrastructure and making it possible to exchange information between different countries securely. QCI will combine terrestrial and satellite segments, where the terrestrial one will use the existing fiber communication networks linking the strategic sites within and between countries, the satellite segment will be deployed to cover very long distances across a large area.\par
Quantum key distribution will be the first service to run this infrastructure~\cite{rosenthal1978photochemical}. It provides the sender and the recipient of an encrypted message with an intrinsically secure random key in such a way that an attacker cannot eavesdrop or control the system. It will secure important confidential communication even against code-breaking by future quantum computers. It will provide services such as securely sharing information, digital signatures, authentication services, and clock synchronization. This infrastructure would be beneficial for the economy and society and ensure the security of sensitive government information both from earth, sea, and in the space \cite{7828091}.\par
The QC and the QML can be key players in 6G wireless networks. The QC can provide the solutions for the 6G networks in the domain of increased channel capacity, e.g., new multiple access technologies, such as NOMA, RSMA demand very high power on run time for computation of \textcolor{black}{successive interference cancellations (SIC)}. Similarly, QC and QML can have a considerable role in 6G in the field of channel estimation, channel coding~(quantum turbo codes), localization, load balancing, routing, and multiuser transmissions~\cite{qml2}. In the communication network core side, QC and QML can solve complex problems such as multi-object exhaustive search by providing fast and optimum path selection to the data-packets in ad hoc sensor networks and Cloud IoT \cite{qml3}.
\subsection{Blockchain}
Blockchain is bringing the revolution to some of the huge industries such as finance, supply chain management, banking, and international remittance~\cite{hewarole}. The concept of blockchain is opening new avenues to conduct businesses. Blockchain provides trust, transparency, security, autonomy among all the participating individuals in the network~\cite{blockchain01}. As far as the telecommunication industry is concerned, innovation in a competitive environment with reduced cost is the most important parameter for the successful businesses in the telecommunication industry. The blockchain industry can benefit the telecommunication industry in the following various aspects.

\subsubsection{\it Internal Network Operations}
Smart contracts in Ethereum, which is the second generation of blockchain technologies, have revolutionized the automation system in various applications. Smart contracts allow the computer code to automatically execute when a certain event is triggered. Because of this fact, blockchain has an immense attraction for its applications in the telecommunication industry to automate various operations such as billing, supply chain management, and roaming. Blockchain can prevent the fraudulent traffic in the telecommunication network thereby saving a lot of bandwidth and resources and ultimately increasing the revenue of the operators~\cite{8726067}. Blockchain can save time for the telecommunication industries and reducing the cumbersome post-billing audit process applying the smart contracts for the authentication and clearance of the bills. Through this process, telecommunication industries can automate accounting and auditing processes.  
\subsubsection{\it Blockchain-based Digital Services}
Telecommunication operators can generate new revenues by proving customers with new blockchain-based services such as mobile games, digital asset transactions, music, payments, and other services. Telecommunication industries can also generate some new revenue streams by allowing the customers to transfer money from the user to user~\cite{le2019prototype}. 
\subsubsection{\it Digital Identity Verification} 
Digital identity verification already costs the government millions of dollars every year. A blockchain-based digital identity verification system can be implemented in the next generation of communication networks which will replace the existing identity verification systems~\cite{le2019prototype}. 

\subsubsection{\it Ecosystem for Efficient Cooperation} 
Next-generation wireless systems aim to provide a variety of new digital services. Blockchain is an attractive application for the complex transactions initiated for these services. Blockchain can also be used in the advertising industry by using user information effectively~\cite{saad2019vision}. This will trigger a massive machine-to-machine~(M2M) transactions. Telecommunication operators can take the initiative of using blockchain in this specific area and usher the next generation of digital services.\par
The demand for massive connectivity in 6G has triggered the network resource management such as power distribution, spectrum sharing, computational resources distribution as the main challenges~\cite{bc2,bc3}. Blockchain can provide solutions to the 6G network in these domains by managing the relationship between operators and users with the application of smart contracts. Similarly, blockchain can solve the unlicensed spectrum management and energy management problems. Blockchain can also be used in seamless environmental protection and monitoring, smart healthcare, cyber-crime rate reduction~\cite{bc5,cloud6}.

\subsection{Tactile Internet}
With the evolution of mobile Internet, sharing of data, videos are enabled on mobile devices. The next stage is the evolution of IoTs, in which communication between smart devices is enabled. Tactile Internet is the next evolution of the Internet of networks, which integrated the real-time interaction of M2M and human-to-machine~(H2M) communication by adding a new dimension of haptic sensations and tactile to this field. Tactile Internet is the term used for transmission of touch over a long distance. Some of the researchers termed it “Internet of Senses”~\cite{6g2020}.\par
ITU has termed the Tactile Internet as the Internet of networks with very high performance, ultra-low latency, high reliability, and high security. Tactile Internet will allow the human and the machines to communicate in the real-time with the environment in a certain range. Haptic interactions will be enabled through Tactile Internet. \par
Creating pressure against the skin without any physical object is one of the main challenges for Tactile Internet. One of the methods to produce such a sense of touch is by intense pressured sound waves. Ultrahaptics, a British company, is working on producing the haptic sensation by using ultrasounds~\cite{simsek20165g}. The ultrasonic transducers can create the sense of touch by controlled production of ultrasonic waves by multiple transducers. These transducers integrated with in-depth cameras can detect the position of the body to react accordingly. Microsoft is also working on the development of haptic sensation using air vortex rings, which are resembling speaker diaphragm~\cite{dohler2017internet}. The concentrated waves from the tiny holes can move with a resolution of 4 inches and to the distance of 8.2 feet~\cite{maier2016tactile}, which has a greater range and much less precise than the ultrasound system.

\subsection{Free Duplexing and Spectrum Sharing~(FDSS)}
In previous wireless generations, wireless systems were using either fixed duplexing~(TDD/FDD) such as in the case of 1G, 2G, 3G, and 4G or flexible duplexing in the case of 5G~\cite{fd1,fd2,fd3}. Whereas, with the progress in the development of duplexing technologies, 6G is expected to use full free duplex in which all users are allowed to use complete resources simultaneously. \textcolor{black}{Users can use all resources (i.e., space, time, and frequency) in a free duplex mode that eventually improves latency and throughput.} \par
Presently, government bodies are monitoring the spectrum and allocating the spectrum to the operators. The owner of the spectrum has the full right to use that spectrum. Any other operator cannot use the spectrum allocated to some other operator. This is only due to the non-development of efficient spectrum monitoring or spectrum managing techniques at present. \textcolor{black}{Therefore, as AI and blockchain are anticipated as key technologies in 6G, robust spectrum monitoring and spectrum management strategies are expected to be developed for the 6G roll-out.} The network resources can be dynamically controlled by AI-aided 6G systems. Therefore, free spectrum sharing will become a reality in 6G.\par
\textcolor{black}{In the context of free spectrum sharing techniques, NOMA is proposed to be a promising multiple access candidate for  B5G/6G communication systems. In NOMA, a complete resource block (frequency band and time slot) is assigned to all users simultaneously, whereas the users are distributed in the power domain. The weakest user receives the maximum power from the BS, whereas the strong users apply the successive interference cancellation (SIC) to the composite NOMA signal to cancel out the messages of the weak users and finally extract their own messages. However, the number of SIC increases exponentially with the increase in the number of users, which increases the complexity of the NOMA system. User cooperation in NOMA can be used to alleviate outage problems of weak users and to provide diversity at the expense of more time slots. However, the number of SICs even becomes larger with the number of cooperating time slots. Space-time block coding-based NOMA (STBC-NOMA) is proposed as an alternative to reduce the number of time slots while keeping the same diversity order \cite{8472261}.} \par
\begin{figure*}[t]
   \subfloat[\label{NOMA1}]{%
      \includegraphics[ width=0.32\textwidth, height=4cm]{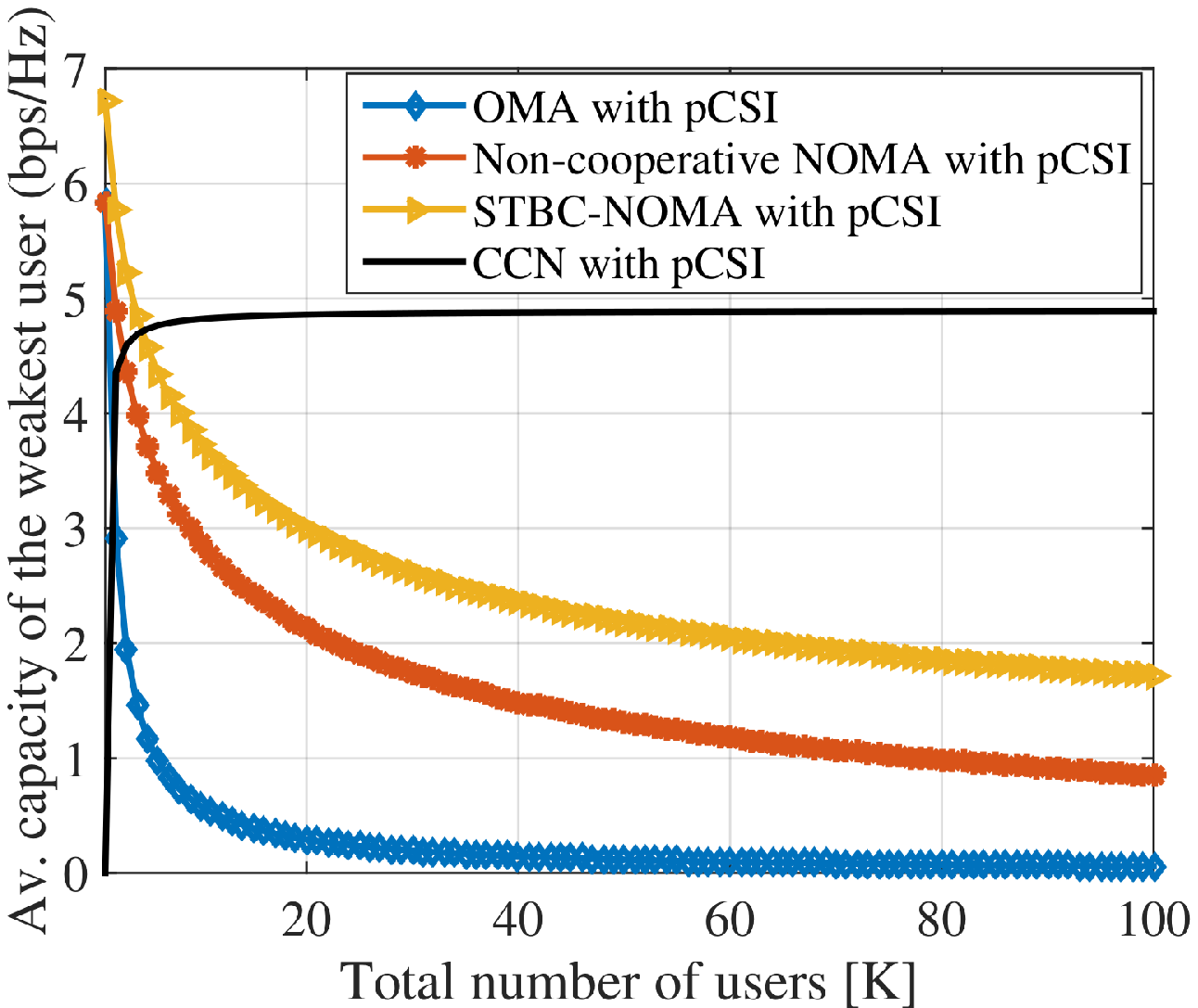}}
\hspace{\fill}
   \subfloat[\label{NOMA2} ]{%
      \includegraphics[ width=0.32\textwidth, height=4cm]{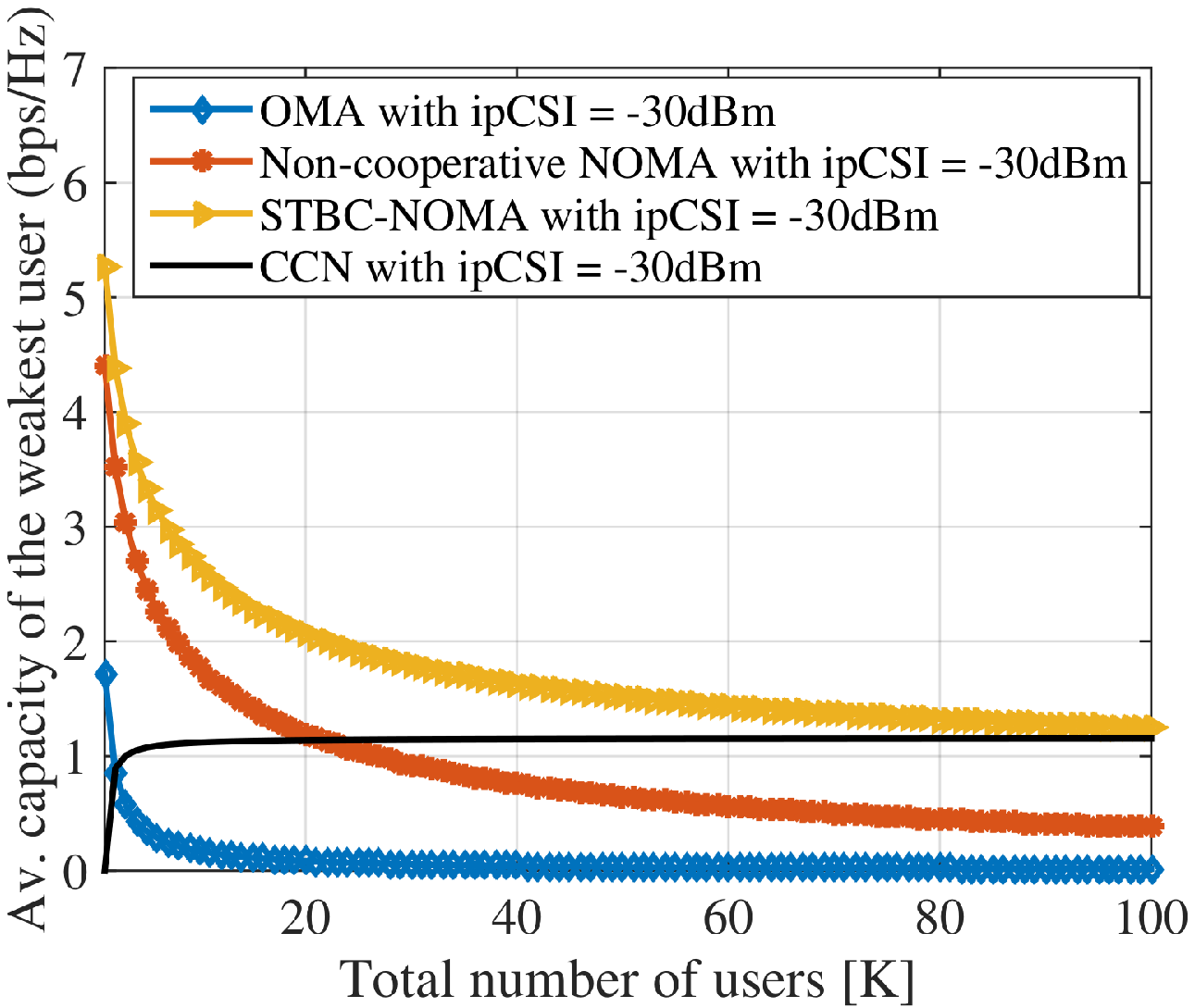}}
\hspace{\fill}
   \subfloat[\label{NOMA3}]{%
      \includegraphics[ width=0.32\textwidth, height=4cm]{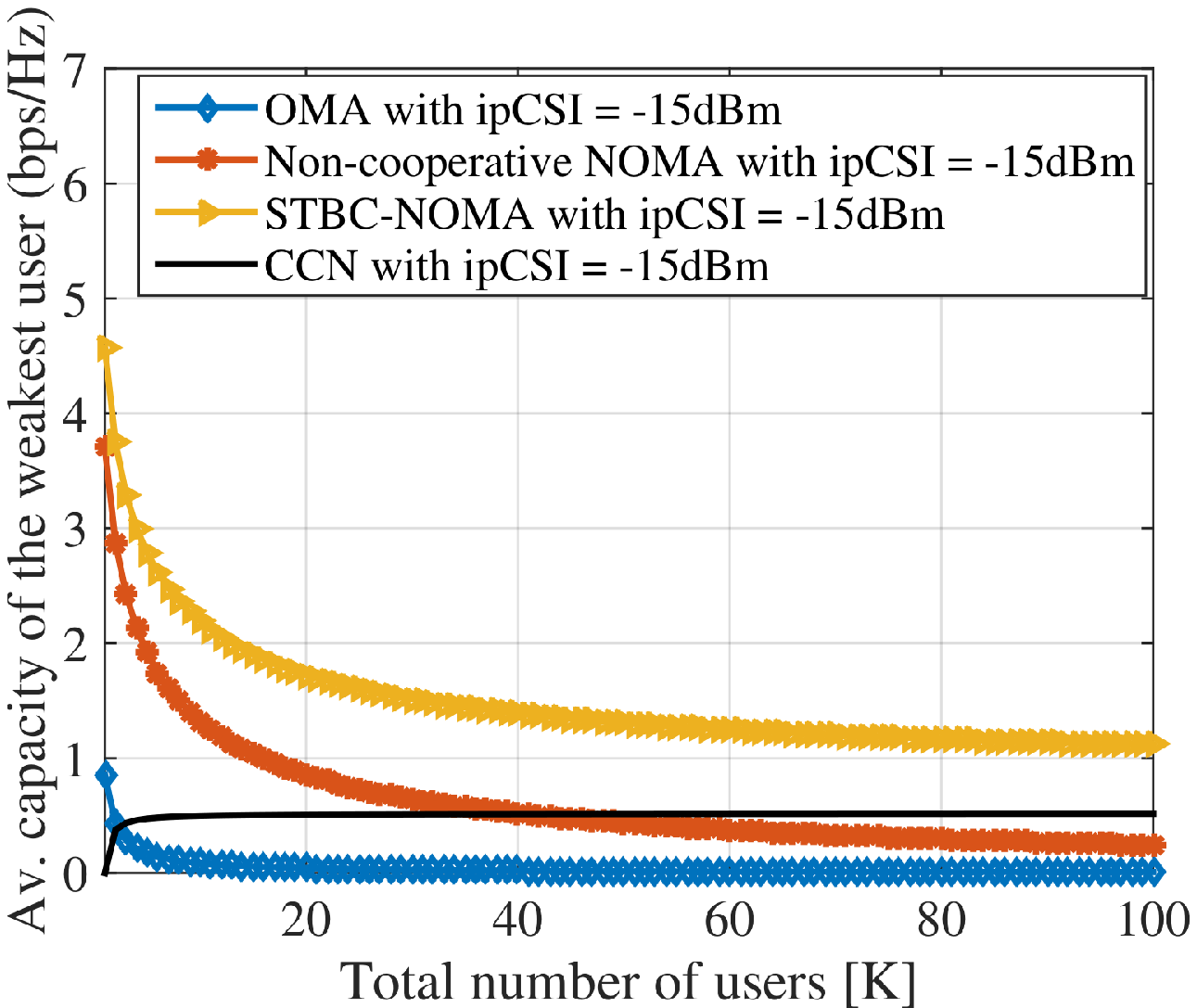}}\\
\caption{\label{workflow}\textcolor{black}{The average capacity vs. total number of users for OMA, non-cooperative NOMA, STBC-NOMA, and CCN with (a) pCSI (b) ipCSI = -30 dBm (c) ipCSI = -15 dBm.}}
\end{figure*}

\textcolor{black}{Apart from imperfect SIC, the imperfection in the channel state information (CSI) also affects the performance of NOMA systems. We present a comparative analysis of the impact of imperfect CSI on the performance of non-cooperative NOMA, conventional cooperative NOMA (CCN), STBC-aided cooperative NOMA, and conventional orthogonal multiple access (OMA) schemes in Fig.~\ref{workflow}. Fig. \ref{workflow} shows the average capacity of the weakest user vs. the total number of users for OMA, non-cooperative NOMA, STBC-NOMA~\cite{jamal2017new} \cite{9170867}, and conventional cooperative NOMA (CCN)\cite{ZDCNOMA} schemes. For a fair comparison between all schemes, we use the same total power budget for all of them. The channel from BS to the users and between users is considered as flat-fading Rayleigh channel. Fig. \ref{workflow}(b) and Fig. \ref{workflow}(c) depict that with perfect channel state information (pCSI), CCN outperforms OMA, non-cooperative NOMA, and STBC-NOMA schemes. However, with the impairments of imperfect CSI (ipCSI), the performance of CCN is severely degraded, where the impact of ipCSI on the STBC-NOMA is much lesser than that of CCN. As shown in Fig. \ref{workflow}(c), with the ipCSI = -15 dBm, the STBC-NOMA outperforms the CCN scheme, whereas the impact of ipCSI on the OMA scheme is negligible. These schemes can be further explored in the future for providing massive connectivity with band-limited applications.} 
\section{6G Application}
Modern society is influenced by intelligent and smart machines. These machines can communicate with each other and with human beings. They can be utilized to help human-being in various aspects of life ranging from medical/e-health, transport, food industry, agriculture, education, etc. In this section, we will introduce some of the important use cases of 6G that will utilize the ideas mentioned above.
\subsection{E-Health and Digital/Bio Sensing} Recently, a widespread pandemic of coronaviruses, a class of viruses that cause sickness in human and animal species, has emerged as a serious threat to the health and life of mankind. The COVID-19 is linked to the family of Severe Acute Respiratory Syndrome~(SARS) which affects the respiratory system of humans and animals~\cite{sars}. With the growing number of COVID-19 infections, there is a requirement for the development of biosensors that are precise, accurate, sensitive, easy-to-use, and specific to detect and monitor infectious diseases. With the development of 6G, these biosensors can be integrated into the smartphones to give an early warning and control the pandemics~\cite{biosens1,biosens2,biosens3,biosens4,biosens5}.\par
\textcolor{black}{With the integration of QC, ML, and biotechnology, 6G networks can be capable of effectively detecting viral diseases by observing the body temperature of infected individuals efficiently. Optical biosensors may also be used to track the pathological functioning of biorecognition substances, such as antibodies, enzymes, whole cells, and DNAzymes, to better detect multiple diseases~\cite{biosens4}. In other areas of electronic health (e-health) such as control of environmental conditions (e.g., temperature, percentage of gases, and light condition), 6G can also be helpful. In various health operations such as emergency care, medical checkups, cleaning contaminated floors, and the supply of medication in rural areas, autonomous robotics can be used. }

\subsection{Holographic Transmissions} 
\textcolor{black}{Holography is a technique for capturing an object 's full 3D image. The methodology was suggested in 1947 by Gabor~\cite{johnston2008cultural}. The term holography derives from the Greek words 'Holo' implies 'complete' and 'graphic' implies 'writing'. An ordinary photograph records the picture's two-dimensional image because it records only the distribution of amplitude or intensity. Therefore, in holography, both the intensity and the phase of light waves are recorded. 
Physical light effects such as interference, reflection , refraction, and diffraction was recorded in holography, and the archive is called the hologram ~\cite{li2018network}. It is possible to play each hologram repeatedly. Although a hologram does not have an object resemblance, it has all the object information in optical forms. Just as mobile cameras have replaced still cameras, video calling and video recording, such as movies, will be replaced by holography.}

\subsection{Communication in Space and Deep Sea}
Space tourism has the immense potential for the next decade both in economic and scientific perspective~\cite{scott2012space}. Humans from every aspect of life will be traveling to space. Several firms are planning to launch sub-orbital commercial flights for space tourism. After some successful and profitable space launches, the next step will be to ensure the availability of space hotels and space hospitals for the customers~\cite{henbest2013private}. Apart from commercial space flights, space research is another potential application~\cite{moro2014new}.\par
6G will expand the range of activities throughout the globe with the availability of easy and effective tools of communication. Autonomous and intelligent robots will be placed in the harsh environmental areas for communication and research purposes~\cite{deepsea}.  The mysteries of the globe could be solved with the aid of powerful and enormous capabilities of 6G networks. Deep-sea exploration such as oil exploration and mineral exploration can become a reality. 
\subsection{Robotics and Automated Vehicles for Beyond Industry 4.0 Era} Industry 4.0 is the term used for the fourth industrial revolution~\cite{ind1}. Industry 4.0 factories have fully automated machines that can self-organize and self-optimize and include processes such as cloud computing, NFV, slicing, and industrial IoT. 6G will bring a new industrial revolution termed as beyond the Industrial 4.0 era. \par 
Robots and fully automated vehicles will take part in the real-time diagnostics, operations, monitoring, and maintenance processes in a very efficient and cost-effective manner ~\cite{ind1,ind2,ind3}. Extremely high reliability and self-organizing feature of automation will come into all aspects of daily life. Swarms of UAVs, through advanced hardware, ML, and QML algorithms, will be used in various operations such as fire control, construction, emergency first response, and agriculture. 
\iffalse
\subsection{IoBT} To be written
\fi
 \section{Key Performance Indicators~(KPIs)}
In this section, we discuss the main KPIs of 6G wireless systems. These KPIs include peak data rate, mobility requirements, connected devices per Km$^2$, area traffic capacity, latency, reliability, network spectral, and energy efficiency.
\subsection{Peak Data Rate}
One of the use cases for next-generation wireless communication is eMBB, which simply implies high data rates. Hence we can download HD videos in a few seconds. The data rate requirements of users are increasing since the birth of wireless communications. As shown in Fig.~\ref{ techn_comp }, 1G had the data rates of a few kbps, which further increased to a few Gbps in 5G. These data rates are still not enough for some applications. Therefore, we require the development of some standard and communication protocols that have data rates in the range of 10-100Gbps~\cite{sliwa2020towards}. \subsection{Mobility}
More mobility robustness is also required in next-generation communication systems. High data rates should be maintained in highly mobile devices. For instance, if we are moving in the airplanes or high-speed bullet trains, the communication should not be disturbed, and data rates should be maintained. The mobility requirements for 6G, as defined by ITU, is $>$1000Km/hr~\cite{sliwa2020towards}.  
\subsection{Massive Connectivity~(Devices/ Km $^2$)}
Another use case for next-generation wireless communication is mMTC. This is the domain where the IoTs comes in and is machine type communication without the interaction of human beings. The calls, messages, and commands are from machine to the other machine. The actions are not carried out by a human. Rather, it is the machines that are communicating with each other. Next-generation wireless networks technology is expected to accommodate $10^7$ devices/Km$^2$~\cite{series2015imt}. \par
Sensor networks and IoTs will be connected to each other in a cooperative way and with several BS. Devices and applications in this category include wearable devices, control and monitoring devices, self-driving cars, smart grids, industrial automation and control devices, and medical and health-related devices. The communication between these devices may be through peer-to-peer or cooperative multi-hop relay manner. \par 
Different applications or devices require different network infrastructure design which could manage different content-driven applications/network. Therefore, keeping all of these requirements in view, next-generation wireless networks would require a completely different approach for planning and optimization.  
\subsection{Area Traffic Capacity~(Mbps/m $^2$)}
With the increase in the number of connected devices per unit area, demand for the higher capacity channels and back-hauling also increased. A highly dense deployed sensor network produces more than tara bytes~(TB) of data on daily basis ~\cite{sliwa2020towards}. This data production needs a high capacity back-hauling channel to accommodate the traffic.\par

In the previous wireless generations~(1G-to-5G), wireless protocols are designed for some specific applications. However, with the development of massive IoTs or mMTC, we need to have some power-efficient and cost-efficient devices to be designed. This massive IoTs communication leads to the development of vehicular communication such as autonomous driving termed as V2X~(vehicle-to-infrastructure). The vehicle needs to interact with another vehicle, with pedestrians, and many other sensors installed in the vehicle. All these communication needs to be extremely reliable and with low latency and security. Industrial automation is another example where a lot of sensors are communicating and generating a huge amount of data. The minimum area traffic capacity limit for 6G is 1000Mbps/m$^2$~\cite{sliwa2020towards}.
\subsection{Extremely-High Reliability and Low Latency with Security~(eRLLCS)}
Low latency means quick and fast communication. We want our packets to be transmitted in a very short amount of time and there should not be much processing delays. The maximum allowable latency in 6G is 10$\mu$ sec~\cite{series2015imt,sliwa2020towards}. The future network of intelligent mobiles and robots will require high reliability and ultra-low latency. Future cities will comprise of smart homes, smart cars, smart industries, smart schools/universities, and smart industries. Smart cities will need to be connected to airplanes, ships, bullet trains, and UAVs. Some of the critical applications which include health care, defense sector, monitoring, and surveillance will require ultra-reliability and low delay.\par
Online gaming services demand high reliability and low latency.
The eRLLCS in 6G wireless systems will integrate the security features with mMTC and URLLC in 5G with greater requirements of reliability of higher than 99.9999999$\%$ (Nine 9's) ~\cite{sliwa2020towards}. Autonomous vehicles will be connected to each other and the communication between them should be ultra-reliable, otherwise it may lead to the loss of lives in accidents. In 6G systems, a lot of households and other sensors will be communicating with each other also require ultra-reliability to prevent any mishap to occur. 
%======Begin Table 1================%
\begin{table*}[t]
\caption{A critical analyses of different techniques proposed for B5G/6G systems.}
\label{table1}
\centering 
\begin{tabular}{ |p{2.5cm}||p{2.5cm}|p{2.5cm}|p{2.5cm}| p{4.3cm}| }

 \hline
 
 %\multicolumn{5}{c}{Table 1: A critical analyses of different techniques proposed for B5G/6G systems.} \\
 \hline
\textbf{Technology Enabler}& \textbf{Pros} &\textbf{Cons} &\textbf{Use Cases} & \textbf{Research Initiatives} \\
\hline

\hline
\hline

\hline
 \textbf{Quantum Communication~(QC) and QML} \cite{nawaz2019quantum,qml1,qml2}&\begin{itemize} 
 \item {Faster}\item {High-performance processing Power} \end{itemize} & \begin{itemize} 
 \item {Costly}\item {Complex} \end{itemize}&\begin{itemize} 
 \item { Drug industry} \item { Radar industry}\item {  Mathematics}
 \end{itemize}& 
 \begin{itemize}
\item { D-Wave Systems Inc.} \item {  IBM Corporation} \item {  Intel Corporation} \item {  Cambridge Quantum Computing Limited} \end{itemize}\\
 \hline
 \textbf{Blockchain} \cite{blockchain01,bc2,bc3,bc5} & \begin{itemize} 
 \item {Distributed}\item{Stability}\item{Integrity}\item{ Immutability}\item{Traceability}\end{itemize}&
 \begin{itemize} 
 \item{Inefficient}\item{High storage}\item{Privacy concerns}\item {Decentralize}\end{itemize} & \begin{itemize} \item { Supply-chain} \item {Voting} \item {Healthcare} \item {Security} \item {Digital identity}   \end{itemize}
 & \begin{itemize}
\item {IBM} \item {Alibaba Group~(China)} \item { Fujitsu ~(Japan)} \item {  Mastercard} \item {  ING Groep~(Dutch banking firm)}\end{itemize}\\
 \hline
 \textbf{Reconfigurable Intelligent Surfaces~(RIS)s} \cite{ris1,ris5} &
\begin{itemize}  \item{Low complexity}\item{Power efficient}\item{Low cost}\end{itemize}
 
 &\begin{itemize}\item{ Difficulty in phase configuration }\end{itemize}& \begin{itemize} \item { Comm. and Defense industry}\end{itemize}& \begin{itemize}
\item { World-wide}\end{itemize} \\
 \hline
 \textbf{Tactile Internet and Haptics} \cite{maier2016tactile,liu2018mobile,dohler2017internet} & 
 \begin{itemize}  \item{Online gaming}\item{Revolutionize the life of disabled peoples}\end{itemize}
 
 & \begin{itemize}\item{Noise  }\item{Less energy efficient  }\item{ Costly }\end{itemize} &\begin{itemize} \item {Mining}\item { Automation}\item { Online gaming}\item {VR and AR} \end{itemize}& \begin{itemize}
\item {TU of Dresden in Germany} \item {Kings College London} \item {Ericsson  and  Microsoft} \item {Ultrahaptics UK}\end{itemize}\\
 \hline
 \textbf{Access Technologies} \cite{ZDCNOMA,rsma1,rsma2,rsma3}& 
 
 \begin{itemize}  \item{Improved network efficiency}\end{itemize}

 & \begin{itemize} \item {Economic Loss due to replacement of existing equipment} \end{itemize} &   \begin{itemize} \item {Variable use cases} \end{itemize} & \begin{itemize}
\item {World-wide}\end{itemize} \\
    \hline
\textbf{New Spectrum \begin{itemize}   \item{ mmWaves} \item{ THz} \item{ VLC} \end{itemize}} \cite{vlc1,vlc2,mmwave4,mmwave3,mmwave2,mmwave1}& 

\begin{itemize}  \item{Higher bandwidth availability}\end{itemize}

 &\begin{itemize} \item {Low penetration power} \end{itemize} & \begin{itemize} \item {Variable use cases} \end{itemize} & \begin{itemize}
\item {World-wide} \end{itemize}\\
  \hline
  %%=====iffalse====
  \iffalse
 \textbf{Cognitive Radios} \cite{coex4,coex3,coex2} &
 
 \begin{itemize}  \item{Low Cost}\item{Spectral and energy efficient}\end{itemize}

 & \begin{itemize} \item {Receiver complexity} \end{itemize} & \begin{itemize} \item {Variable use cases} \end{itemize} & \begin{itemize}
\item {World-wide} \end{itemize} \\
  \hline
  \fi
  %%=====fi====
 \textbf{Internet-of-Everything~(IoE)} \cite{cloud5,3gpprel14,aiml4} &  
 
 \begin{itemize}  \item{Low latency}\item{Higher data rates}\end{itemize}
 
  & \begin{itemize} \item {Low energy efficiency} \end{itemize} & \begin{itemize} \item {Variable use cases} \end{itemize} & \begin{itemize} \item {Worldwide}\end{itemize}\\
    %%=====iffalse====
  \iffalse
  \hline
 \textbf{Small cells} \cite{3,holma2016lte,mmwave4} & 
 
 \begin{itemize}  \item{Higher achievable data rates.}\end{itemize}

   & \begin{itemize} \item {More equipment cost } \end{itemize}  &\begin{itemize} \item {Higher density areas}\end{itemize}& \begin{itemize}
\item {World-wide}\end{itemize}\\
 \hline
 \fi%%-----
 \hline
\end{tabular} 
\end{table*}
%======End Table================%
\subsection{\textcolor{black}{A QoS-aware Spectral Efficient Network}}
The future intelligent wireless network will comprise of intelligent/smart factories, intelligent/smart hospitals, schools, universities, and autonomous robots. This will require a highly spectral efficient network having high computing power. A high-density and high-rate network will require high bandwidth. The scarcity of the bandwidth will increase with the increase of data in the network. \textcolor{black}{For reliable communication, an ultra-spectral efficient network would therefore be needed while at the same time satisfying the criteria for QoS of all users in next-generation wireless networks, which will be smart enough to move to a new state with changing environmental conditions.} \par
With the increasing number of mobile devices and communication types, the scarcity of the radio wireless spectrum has increased. Therefore, some communication protocols are needed to be designed for spectral efficient communication. So that the bandwidth resource is effectively utilized. \textcolor{black}{The spectral efficiency of 6G networks is supposed to be $> $15 times that of 3G~\cite{sliwa2020towards}.}

\subsection{Energy Efficient Network}
The next-generation wireless communication system will consist of massive self-organizing and self-healing robots. All these intelligent robots/devices will require high computation power. Therefore, the need for energy will be increasing with the increase in intelligent robots. Currently, traditional GPUs are not meeting the energy efficiency requirements of next-generation wireless networks communication networks. In such a scenario, an energy-efficient and scalable intelligent network design will be required. The industry has moved towards IoTs, IoEV and IoBTs~\cite{Qian,sliwa2020towards}. We have sensors deployed everywhere. There is a sensor in our door, in our air conditioner, in our car, on the TV, in the refrigerator, in offices. All these sensors need energy-efficient communication. \par
Table~\ref{table1} gives a critical analysis of the potential 6G technologies, which will enable communications in the B5G/6G era. Advantages and disadvantages along with the research initiative in these technologies are also described. 

\section{Research Challenges and Directions}
\subsection{ Hardware Complexity and Increase in Chip Size}
Next-generation mobile communication will integrate multiple communication devices ranging from sensors to HD video transmission devices or communication with high-speed trains and airplanes. Such devices will require variable packet sizes to be transmitted, contrary to the previous mobile communication technologies with fixed packet size.\par
With the variable packet size to be transmitted, the hardware complexity is increased. Since the frame/packet size is not known to the mobile station a priori, it must select its signal processing and RF chains according to the incoming packet, which means that the chip size will be increasing, which will ultimately increase the mobile size. This will again increase the processing time which is not desirable in next-generation wireless communication. \par
6G communication network will cover a large bandwidth ranging from 3GHz to 60GHz. Therefore, to enable communication, at any frequency in this band, the complexity of the hardware circuitry is increased. For each, the communication, antenna, RF filters, amplifiers have to be designed accordingly. To cover all the band in next-generation communication systems, multiple RF and signal processing chains would be required, which would ultimately increase the chip size and hardware. \textcolor{black}{Therefore, a lot of focus needs to be paid to the field of open research.} 
\subsection{\textcolor{black}{Variable Radio Resource Allocation}} \textcolor{black}{With variable QoS requirements, a variable radio resource has to be assigned to the user. This may be variable bandwidth, power, or both. Another challenging aspect of 6G is that with propagation frequency, the signal can attenuate rapidly and have high penetration loss at high frequencies. The signal is also automatically attenuated upon accessing houses, residences, or workplaces. With the increasing frequency, as the radio waves get attenuated, it could have trouble penetrating walls in buildings and houses, while affecting the QoS requirements of the users. It is therefore of great significance to develop stable, fast, precise algorithms to manage the 6G communication requirements by dynamically allocating variable resources.} 
\subsection{ Ultra-Low Power Circuits with High-Performance Processing Capabilities} To cope with latency-critical scenarios such as automated cars and medical/health applications, the communication needs to be accomplished in a very short duration of time. To attain the latency of only a few milliseconds is quite challenging. To achieve low latency and ultra-high reliability, it is essential to develop powerful high-end processors that consume low power.%, which is quite challenging. }
\subsection{ Pre-emptive Scheduling in Massive Connectivity} In the wireless communication, a network node assigns the radio resources to other nodes as per their priorities~\cite{preemp}. However, in the case of 6G wireless networks, where a massive number of devices will be connected to each other, setting the priority level for all these devices and maintaining the latency and packet loss requirements of 6G will be quite challenging. \textcolor{black}{In order to carry out such pre-empted communication on the 6G wireless network, which is another open research field, some appropriate algorithms would be required.} 
\subsection{Seamless Coexistence of Multiple RATs}\textcolor{black}{With multiple interfaces, 6G can incorporate multiple RATs. This ensures that 6G will coexist with and interoperate with other technologies, such as Wi-Fi, Bluetooth, networks of ad hoc sensors, and IoTs, etc. For the change in the user's radio environment, the air interface can dynamically change. It is still an open challenge to develop scalable techniques that guarantee interoperability while meeting the KPI requirements of 6G.}  
%\pagebreak
\subsection{Security and Privacy}\textcolor{black}{In the last few years, the number of IoT devices has grown exponentially. These devices include personal IoT, health care IoT, and industrial IoT, which are connected to form a mesh network. 6G is expected to be an enabler for large scale cyber operations including IoT applications. As IoT devices are connected to the internet, broad-scale distributed denial of service (DDoS) attacks could be more common. This large-scale DDoS attack will serve as an enabler in a 6G IoT system that can lead to security, privacy, and trust issues in the network. This is therefore an open research challenge for 6G networks, too.}
\section{Conclusion}
Technology has a great impact on the lifestyle of human beings. Wireless technologies have revolutionized businesses, living standards, infrastructure, and many other aspects of human life. Mankind is in a constant struggle to find elegant solutions to various problems and is in search of new avenues to progress. This desire of mankind has evolved wireless communication from 1G to 5G. However, this development has not stopped here. The researchers around the world are working hard for the development of 6G communication network, which is expected to be rolled out by 2030.\par
In this paper, we covered various aspects of 6G wireless networks with different perspectives. We provided a vision for B5G/6G communications, 6G network architecture, KPI requirements, key enabling technologies, their use-cases, and network dimensions that will landmark the next generation communication systems. Furthermore, a way out is discussed how these potential technologies will meet the KPI requirements for these systems. \textcolor{black}{Finally, the opportunities and research challenges such as hardware complexity, variable radio resource allocation, pre-emptive scheduling, power efficiency, coexistence of multiple RATs, and security, privacy and trust issues for these technologies on the way to the commercialization of next-generation communication networks are presented.}

\section{declarations} 
\subsubsection{Availability of data and material}
Not applicable
\subsubsection{Competing Interest}
The authors declare that they have no competing interests.
\subsubsection{Funding}
The authors extend their appreciation to the Researchers Supporting Project number (RSP-2020/32), King Saud University, Riyadh, Saudi Arabia for funding this work.
\subsubsection{Authors' contributions} 
MWA, SAH, RG suggested and conceived the core conception of this research work. RG presented the case study, findings and discussion. SG, MSH and HJ defined the overall organization of the manuscript. MSH  has carried out thorough oversight of this work. The final manuscript was read and accepted by all contributors.
\subsubsection{Acknowledgements}
Not applicable
%\bibliographystyle{unsrtnat}
%\bibliographystyle{cas-model2-names}
%\bibliography{sample}
 \bibliographystyle{splncs03_unsrt}
\end{document}